\documentstyle[12pt]{article}

\newcommand{\undtilb}{\,\mathrel{\lower9pt\hbox{$\tilde{}$}}
\hskip-5.5pt \hbox{$b$}}

\newcommand{\undtild}{\,\mathrel{\lower9pt\hbox{$\tilde{}$}}
\hskip-5.5pt \hbox{$d$}}

\newcommand{\undtilk}{\,\mathrel{\lower9pt\hbox{$\tilde{}$}}
\hskip-5.5pt \hbox{$k$}}

\newcommand{\undtildel}{\,\,\mathrel{\lower9pt\hbox{$\tilde{}$}}
\hskip-7.5pt \hbox{$\Delta$}}

\newcommand{\undtildelb}{\,\,\mathrel{\lower9pt\hbox{$\tilde{}$}}
\hskip-4.5pt \hbox{$\Delta$}}

\newcommand{\lsim}{\mathrel{\lower4pt\hbox{$\sim$}}
\hskip-12.5pt\raise1.6pt\hbox{$<$}\;}

\newcommand{\gsim}{\mathrel{\lower4pt\hbox{$\sim$}}
undtilb\hskip-12.5pt\raise1.6pt\hbox{$>$}\;}

\begin{document}
\title{Correlated Scattering and Cluster Planes}

\author{Shmuel Nussinov \\
{\it School of Physics and Astronomy} \\
{\it Raymond and Beverly Sackler Faculty of Exact Sciences} \\
{\it Tel Aviv University} \\
{\it 69978 Tel Aviv, Israel} \\
and \\
{\it Brookhaven National Laboratory} \\
{\it Physics Department} \\
{\it Upton, NY\ \ 11973\footnote{E-mail: Nussinov@ccsg.tau.ac.is}}}

\maketitle
\begin{abstract}We suggest a new correlation in diffractive production
of 2 ``clusters'' $A+B\to A^\ast B^\ast$ with large intrinsic angular
momenta for each $A^\ast$ and $B^\ast$ cluster. These correlations are
expected in the context of the ``color dipole picture'' for high energy
collision and reflect the approximate conservation of dipole direction
during the collision. This conservation is in particular manifest when
the two dipoles, $\vec d_A$, $\vec d_B$ and the impact vector
$\undtilb$ are all parallel. The predicted positive triple correlation
is between the momentum transfer $\undtildel$ and the planes of the
$A^\ast$ and $B^\ast$ clusters.
\end{abstract}

QCD is well-tested in the short distance perturbative regime, yet it
cannot predict the bulk of hadronic scattering data without additional
``models'' and approximations.

Some general features of the energy dependence of cross-sections
\cite{low}, and helicity conservation  at high energies,
trace back directly to the vectorial (spin1) nature of the gluon. The
interaction with opposite sign of the latter with the quark
(anti-quark) in a meson underlies the ``color dipole model'' of hadrons
\cite{chen}. 

Our purpose in the following is to use the latter model and variants
thereof to motivate yet another general feature of hadronic collisions,
namely the tendency of the scattering plane and the planes of the
final state particle clusters to be parallel.

During the short $\{$Lorentz contracted$\}$ time of a high energy
collision the $q\bar q$ in, say the projectile meson, are relatively
frozen and we have an instantaneous, well defined dipole $\vec d$.

This observation underlies the ``Color transparency'' phenomenon
\cite{frank} --- associated with hadrons which enter the collision as
small dipoles. While such hadrons can scatter effectively at large
angles --- in say the center of a large nucleus, the latter is almost
transparent to these ``point-like configurations'', which can traverse
it unscathed and cause no excitation \cite{farrar}.

The effect we consider here is due to the direction of the color
dipoles which also tends to be conserved\cite{nuss}.

Thus let us assume that the colliding hadrons have instantaneous
dipoles $\vec d_A$, $\vec d_B$, and that at the time of collision the
centers of the dipoles (Hadrons) (which move along the $\pm z$
directions), are separated by a transverse (impact parameter)
$\undtilb$. [We use $\tilde{}$ to indicate the transverse part]. We
will argue that i) the scattering will be enhanced when $\vec d_A$,
$\vec d_B$ and $\undtilb$ are parallel and ii) the hadrons in the
final states will tend to align along this common conserved direction.

We will next proceed to discuss this in a variety of settings ---
perturbative, and others. 
\medskip

\noindent (a) The feature we suggest is quite obvious in a deep
inelastic, large momentum transfer $\Delta$, limit. The reaction is
then dominated by one hard gluon exchange between a quark (or
anti-quark) in A and a quark (or anti-quark) in B\null. The scattered
(anti) quarks generate two jets: the ``A jet'' from A and the ``B jet''
from B\null. Up to corrections due to intrinsic $P_T$ and color
neutralization we expect --- on essentially kinematic grounds --- that
both jets will align with each other. However, in this case we do not
have a third independent direction in addition to that of the two jets.
Rather we may have the ``wounded'' forward and backward hadrons which,
having each lost an (anti) quark continue, {\it essentially\/} along
the initial direction. The non-trivial and not purely kinematical
prediction we make here is that the slight deviation of the ``wounded
Hadron'' forward (A-q) and the backward (B-q) jets from their initial
$\hat z$ directions will be --- due to the pull of the kicked quark ---
also along the same plane defined by $\Delta$ and the transverse jets
axis.

It may be difficult however to precisely separate the forward and
backward ``wounded'' (A-q and B-q) hadron jets from the rest and verify
this statement. To have a precise determination of $\Delta$ we need in
principle to measure all the particles in the $A^\ast$ {\it or\/} in
the $B^\ast$ cluster. On the other hand the definition of the plane
$A^\ast$ and 
$B^\ast$, just like for the case of transverse jets does not require
that all the particles in the cluster will be measured.

Our suggestion is that some similar alignment feature manifests, albeit
in a weaker fashion, in a large class of hadronic interactions ---
where much more that kinematics is required to justify it.
\medskip

\noindent(b) The directions of the dipoles are not always manifest.
Thus let us consider an elastic scattering of spinless hadrons $A+B\to
A+B$. the collision $(\hat z,\undtildel)$ plane is clearly
well-defined and corresponds to the impact $(\hat z,\undtilb)$
plane. To see this assume we generate by an appropriate superposition
an initial relative A-B state aligned along the $\phi=0$, $\hat x$
axis: 

\begin{equation}
|\Psi_i \rangle=\sum_l C_l \sum^l_{m=-1} |\Psi_{lm}\rangle |Y^m_l
(\theta,\phi)    \label{eq1}
\end{equation}

After the collision we will have:

\begin{equation}
| \Psi_{f_{scatt}} \rangle= \sum_l C_l (e^{2i\delta_l}-1)
\sum^{l}_{m=-1} |\Psi_{lm} \rangle |Y^m_l (\theta,\phi)   
\label{eq2}
\end{equation}

\noindent with the $m$ independent phase shifts $\delta_e$.
Hence also $|\Psi_{f_{scatt}}\rangle$ is aligned along the same $\phi=0$
plane. Thus the scattering from the central potential conserves the
initial collision plane, as expected.

Unfortunately the projection onto the spinless initial and final
particles erases any information regarding the instantaneous  dipole
directions and no vector is available to correlate $\undtildel$ with.

\medskip

\noindent (c) Let us next consider diffractive scattering where one or
both of the colliding particles transforms into ``clusters of
fragmentation products'': $A+B\to A^\ast+B^\ast$. The distinguishing
characteristic of diffractive scattering is the existence of a
``rapidity gap''. It clearly separates the particles in the $A^\ast$
cluster from those in the $B^\ast$ cluster. Also since a color-flavor
singlet system (the ``Pomeron'') is exchanged between the $A$-$A^\ast$
and $B$-$B^\ast$ vertices, the clusters maintain the initial valence
quark flavors. In the ``color dipole model'' the reaction can then be
viewed as the scattering of the two bound states $A=q_a\bar q_a$,and
$B=q_b\bar q_b$, into the excited state $A^\ast$, $B^\ast$. If the
latter have sufficiently high spins $S^\ast_A$, $S^\ast_B$ and hence
also large orbital angular momenta between $q_a\bar q_a$ in the
$A^\ast$ rest frame, (and likewise for the relative motion of $q_b\bar
q_b$, inside the $B^\ast$ cluster), then the planes of the clusters can
be well-defined. Indeed we could construct in analogy with
eq.~\ref{eq1} ``aligned states'' of $q_a \bar q_{a^\prime}$ say by
superimposing the $2L_A+1$ $m$ states of the relative quark motion. The
plane of the cluster is then defined --- due to $\Delta
L\Delta\phi\approx 1$ uncertainties --- to within $\Delta \phi \approx
\frac{1}{L_{A^\ast}} \approx \frac{1}{S^\ast_A}$. If the $A^\ast$
decays into two spinless hadrons then the axis of the $A^\ast$ cluster
is identified with the relative decay direction of these two particles
in the $A^\ast$ rest frame. If $A^\ast$ decays into several hadrons we
can reconstruct the $A^\ast$ axis using appropriate ``collective
variables'' --- similar to those used in jet analyses. We will restrict
ourselves to $\Delta << P_z$ so that while the overall scattering
$\undtildel$, $P\hat z$ plane of the $A+B\to A^\ast+B^\ast$ reaction is
well-defined, the Lorentz transformation from the overall center mass
frame to the $A^\ast$ (or $B^\ast$) rest frame are boosts mainly along
the $z$ axis. We can then compute, for each of the final decay
particles in $A^\ast$, the scalar product $\undtilk_{Ai} \cdot
\undtildel^i, i=1,\dots n_A$. to avoid the kinematical biasing due to
$\sum \undtilk_{Ai} \equiv \undtildel$ we modify it to:

\begin{equation}
f^A_i = \left(\undtilk_{Ai} - \frac{\undtildel}{n_A} \right) \cdot
\undtildel \label{eq3}
\end{equation}

\noindent The claim we are making then is that:

\begin{equation}
|\cos \theta^A_i| = \frac{(f^A_i)}{\left|\undtildel\right|\left|
\left( \undtilk_{Ai} - \frac{\undtildelb}{n_A}\right) \right|} \qquad
i=1,\dots n_A \label{eq4}
\end{equation}

\noindent are {\it not\/} uniformly distributed in the 1--0 interval
but are more concentrated near unity. [And likewise for the
similarly-defined $|\cos\theta j^B|$ $j=1,\dots n_B$.] Let us next
motivate the above dynamical correlation.
\medskip

\noindent (c$_1$) Let us consider first the case when the impact
parameter between the initial spinless particles $b=|\undtilb|$ is
larger than typical hadronic sizes.

In analogy with the Van-der-Waals-interactions in Q.E.D. we can assume
that the initial unpolarized particles produce --- via mutual
interactions --- dipole moments in each other. At say $t=0$, we assume
that the initial colliding particles are at $z=0$, the point of nearest
approach. The mutual forces will at this point be predominantly along
the direction of $\undtilb$, and the induced dipoles $\vec d_A$ and
$\vec d_B$ will also align with it. Conversely let us assume that we
have at $t=0$ some dipoles $\vec d_A$ and $\vec d_B$. The color
analogue of the electromagnetic dipole interaction is approximated to
be: 

\begin{equation}
V_{dd} (E.M.) \approx \frac{\vec d_A\cdot \vec d_B}{r^3} - \frac{3(\vec
d_A\vec r)(\vec d_B\vec r)}{r^5} \label{eq5}
\end{equation}

\noindent with $\vec r =\vec b+z\hat e_z$ the instantaneous separation
between the dipoles. This interaction is clearly maximal if $\vec
d_A,\vec d_B$ and $\undtilb$ are all aligned. The large classical
interactions are reflected quantum mechanically in stronger scattering
probability. Hence we expect a positive correlation between the
directions of $\vec d_A,\vec d_B$ and $\undtilb$. Since the latter is
in the direction of $\undtildel$ we have then a $\vec d_A,\vec d_B$ and
$\undtildel$ positive correlation.
\medskip

\noindent (c$_2$) The tendency of $\vec d_A$ and $\vec d_B$ to be
parallel persists also in the opposite limit i.e.\ when $\undtilb =0$
and the hadrons strongly overlap.

Let us assume that the initial hadrons $A$ and $B$ have similar
chromoelectric fields $\vec E_A(\vec r)$ and $\vec E_B(\vec r)$ with
$\vec r$ referring to the centers of $A$ and $B$, in the first and
second case, respectively.

Note that we do not necessarily assume that $\vec E_A$ and/or $\vec
E_B$ have the perturbative dipole form. Rather we can incorporate
various non-perturbative features by assuming that $\vec E_A$ and $\vec
E_B$ correspond --- say --- in the case of large $q_a\bar q_a$ (or
$q_b\bar q_b$) separation --- to a confined color flux tube
configurations. 

In a perturbative-impulse treatment of the collision process itself we
approximate the interaction by the mutual chromoelectric energy
\cite{nusstwo}. 

\begin{equation}
V_{\rm int} \approx \int\vec E_A(\vec r) \cdot \vec E_B(\vec r) d\vec r
\label{eq6} 
\end{equation}

Clearly the above overlap integral (at $t=0$, $b=0$) is maximized when
the $\vec E_A$ and $\vec E_B$ configurations are ``parallel'' both in
the internal color space (Apposteriori --- justifying the Abelian ---
Q.E.D. like approximation adopted here) and in real space i.e.\ when
the dipoles and/or flux tubes of $A$ and $B$ are parallel.
\medskip

\noindent (c$_3$) Let the quark $q_a$ and anti-quark $\bar q_a$ be in
some instantaneous color field $\vec E$ (due say to the other particle
$B$, ($B^\ast$)). The sum of the forces acting on $q_a$ and $\bar q_a$
imparts the overall momentum transfer, $\Delta$ to the $A^\ast$
cluster. The difference of the forces tends to separate $q_a$ and $\bar
q_a$ from each other and thus to induce the $\vec d_A$ dipole. $\{$For
uniform $\vec E$ fields only the latter effect is operative.$\}$ The
same force difference generates a relative momentum between $q_a$ and
$\bar q_a$ and cause the ensuing stretching and excitation from $m_A$
to $m^\ast_A$. As the quarks $q_a$ and $\bar q_a$ recede from each
other we expect --- from the electric fluxtube model for multi-particle
production \cite{neub} --- (which underlies the Lund model
\cite{ander}) that the hadrons formed via the Schwinger mechanism in
this field, will align along the tubes, i.e.\ the $\vec d_A$ axis.
There is of course a ``transverse'' scatter of order $\Lambda_{QCD}$ relative to
this axis, but, to the extent that $M^\ast_A>> \Lambda_{QCD}$, it will
not destroy the basic feature of $\undtild_A$ and $\undtilk_i$
alignment. This provides then another key element in our chain of
arguments by correlating the axes of the $A^\ast$ cluster with $\vec
d_A$ (and likewise for $B$).

We should emphasize that we need not use specific variants of the color
flux tube (or lund) models. Rather we appeal to the general
``jet-like'' fragmentation of the $q\bar q$ system with limited small
momentum in the directions transverse to the jet axis.

Also note that the $A$-$B$ interactions which generate the dipoles (and
cause also the momentum transfer $\Delta$) need not be perturbative one
gluon exchanges. Rather we could have the dipoles and the relative
separation of the $q\bar q$ --- as well as the overall $\Delta$ --- be
built via many soft ``eikonal-like'' interactions. This corresponds to
solving for the quark propagators in a background gluon field (as is
done in order to derive the Schwinger formula for $q\bar q$ pairs
production). 
\medskip

\noindent (d) Finally we note that whereas mutual torques may in
general tend to rotate the dipoles during the collision this in {\it
not\/} the case in the stationary special case of interest with $\vec
d_a|| \vec d_b|| \undtilb$.

If  the initial particles are
polarized photons or nucleons there  could {\it a priori\/} be  further
correlations 
involving --- in addition to the scattering and ``cluster'' planes ---
also the initial polarization.

Thus let us have a transversely polarized photon diffractively scatter
into a $\rho$ or $\phi$ vector meson which then decay into $\pi^+\pi^-$
or $K^+K^-$. The decay plane does indeed tend to correlate with the
initial plane of polarization. This is however simply explained by
helicity conservation/independence of the high energy process. The
photo-produced $\rho$ or $\phi$ have the same transverse polarization
as the initial photon, and the decay amplitude $\hat \epsilon_V (\vec
k_1-\vec k_2)$ generates final state pseudoscalar with relative
direction $\tilde n$ having a $\cos^2\theta$ distribution relative to
$\hat\epsilon_V$. However the $\rho$, $\omega$, and $\varphi$ are the
$^3S$ analogs of spin $^1S\pi$ and the $\eta\eta^\prime$. Hence the quarks in
$\rho$, $\omega$, $\varphi$ have to lowest order no orbital angular
momentum and we do 
not expect the dipoles to correlate with $\hat\epsilon_V$ or the
scattering plane. To see the effect of interest here we need,
unfortunately, to produce higher $L$ excitations.

We have not made quantitative predictions for the magnitude of the
correlations suggested here. We believe however that the extreme
generality and relative model independence of these make them yet
another most valuable general test of Q.C.D., which can be readily
applied to a wealth of existing and future data. 
\medskip

\noindent Acknowledgement: I have enjoyed discussions with D. Tedeschi
on $\phi$ photoproduction which (indirectly) inspired this work.


\begin{thebibliography}{9}

\bibitem{low} F.E. Low, Phys.\ Rev.\ {\bf D12}, 163 (1975). \\
S. Nussinov, Phys.\ Rev.\ Lett.\ {\bf34}, 1286 (1975) and Phys.\ Rev.\
{\bf D14}, 246 (1976).   

\bibitem{chen} Zhong Chen and A.H. Mueller,  Nucl.\ Phys.\ {\bf B451}, 579--604 (1995).   

\bibitem{frank} L. Frankfurt, G. A. Miller and M. Strikman,  Comm.\
Nucl.\ Part.\ 
Phys.\ {\bf21}, 1--40 (1992).   

\bibitem{farrar} Glenns R. Farrar, Leonid L. Frankfurt and Mark I.
Strikman,  Phys.\ Rev.\
Lett.\ {\bf64}, 2996-2998 (1990).   

\bibitem{nuss} S. Nussinov and J. Szwed, Phys.\ Lett.\ {\bf B84}, 945
(1979).   

\bibitem{nusstwo} S. Nussinov, Phys.\ Rev.\ {\bf D50}, 3167 (1994).  

\bibitem{neub} A. Casher, H. Neuberger and S. Nussinov, Phys.\ Rev.\
{\bf D20}, 179 (1979).   

\bibitem{ander} B. Anderson {\it et al}., Phys.\ Rep.\ {\bf97}, 31 (1983).



\end{thebibliography}
\end{document}